\documentclass{article}

\usepackage{arxiv}

\usepackage[utf8]{inputenc} 
\usepackage[T1]{fontenc}    
\usepackage{hyperref}       
\usepackage{url}            
\usepackage{booktabs}       
\usepackage{amsmath}        
\usepackage{amsfonts}       
\usepackage{nicefrac}       
\usepackage{microtype}      
\usepackage{lipsum}
\usepackage{graphicx}
\usepackage{adjustbox}      
\graphicspath{ {./images/} }
\title{Quantum Sampling Architecture for Protein Structure Reconstruction on Utility-Scale Hardware}

\author{
 Yuqi Zhang \\
  Kent State University\\
  Kent, OH, USA \\
  Cleveland Clinic\\
  Cleveland, OH, USA \\
  \And
 Bo Fang \\
  University of Texas at Arlington\\
  Arlington, TX, USA \\
  \And
 Yuxin Yang \\
  Cleveland Clinic\\
  Cleveland, OH, USA \\
  \And
 Feixiong Cheng \\
  Cleveland Clinic\\
  Cleveland, OH, USA \\
  \And
 Jieyang Chen \\
  University of Oregon\\
  Eugene, OR, USA \\
  \And
Sherry Fang \\
Regailator Inc. \\
Stow, OH, USA \\
  \And
 Siwei Chen \\
  University of Chicago\\
  Chicago, IL, USA \\
  \And
 Junhan Zhao \\
  University of Chicago\\
  Chicago, IL, USA \\
  \And
 Qiang Guan$^{*}$ \\
  Kent State University\\
  Kent, OH, USA \\
  \texttt{qguan@kent.edu}
}

\begin{document}

\maketitle

\begin{abstract}
Predicting the structure of short peptides in protein binding pockets remains difficult because this regime requires physics-based conformational search, yet existing methods do not provide a practical way to carry out that search on current hardware. We present \textit{QSAD}, a quantum-classical framework that reformulates peptide structure prediction as amino-acid-level Hamiltonian sampling and replaces iterative optimization with non-iterative Hamiltonian evolution. Executed entirely on IBM Heron~R2 across 101 binding-pocket peptides (5--18 residues), QSAD improves prediction accuracy by 27--71\% over all evaluated AI and quantum baselines while maintaining the lowest variance across tested lengths. QSAD also tolerates noise levels 3--5$\times$ beyond typical hardware error rates, where iterative methods fail, and reduces mean quantum execution time by $27\times$ relative to VQE. The sampled ensemble further supports approximate reconstruction of protein energy landscapes. These results establish coarse-grained quantum sampling as a practical computational path for structure prediction in regimes where data-driven methods lack sufficient signal.
\end{abstract}

\section{Introduction}

\begin{figure}[htbp]
\centering
\includegraphics[width=0.8\linewidth]{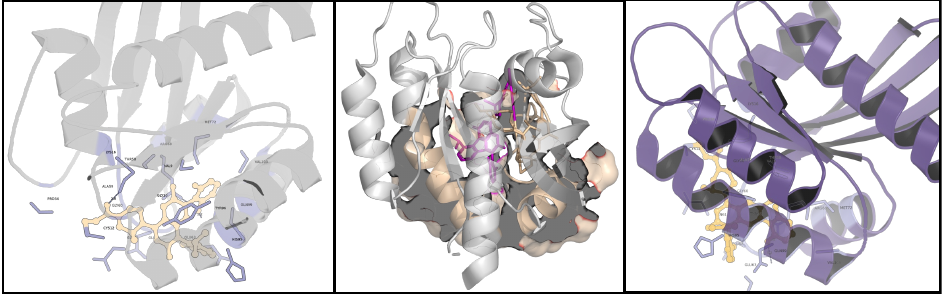}
\caption{Pocket--ligand structure of 6OIM, a KRAS G12C system. This example shows the local binding-pocket region studied in this work.}
\label{fig:6oim_kras_g12c}
\end{figure}

Predicting the three-dimensional structure of a protein from its amino-acid sequence is a central problem in computational biology, with direct importance for understanding biological function, disease mechanisms, and drug design~\cite{yin2024leveraging}. Within this broad problem, one specific regime remains particularly difficult in practice: short peptides of 5--20 residues located in binding-pocket regions. These segments mediate interactions between proteins and drugs, substrates, or partner proteins, so that their geometry directly influences binding affinity and biological activity. They are difficult to predict because protein folding is fundamentally a process of exploring the energy landscape to identify stable conformational basins. However, such systems lack the intuitive sequence and evolutionary patterns that enable modern prediction methods, such as deep learning, to work effectively, and their energy landscapes are substantially more complex than those of protein systems that follow more general structural regularities. As a result, existing methods perform poorly in this regime~\cite{manfredi2024alpha,zhang2025qdockbank}. Figure~\ref{fig:6oim_kras_g12c} shows 6OIM, a representative pocket--ligand structure of the cancer target KRAS G12C.

An effective solution must satisfy two conditions at once. First, it must derive structure by mapping the underlying energy landscape from physical principles, because short peptides do not provide enough sequence signal to support reliable statistical learning. Second, it must search an exponentially large and highly complex energy landscape efficiently enough to recover low-energy structures without exhaustive enumeration.

The conformational search space in binding-pocket regions is extremely complex. Their rich biochemical characteristics give rise to a highly rugged energy landscape, significantly increasing the problem complexity. Classical methods such as molecular dynamics~\cite{swope2004describing} and Monte Carlo sampling~\cite{hansmann1999new} therefore incur very high computational cost, and the search process is highly prone to becoming trapped in local optima~\cite{zhang2015application}. Deep learning methods such as AlphaFold2~\cite{yang2023alphafold2} and AlphaFold3~\cite{abramson2024accurate} rely heavily on informative long-sequence patterns, and thus fail to produce reliable predictions in short-peptide settings where information is limited and structural variability is high~\cite{manfredi2024alpha,zhang2025qdockbank}.

\begin{figure}[htbp]
\centering
\includegraphics[width=\columnwidth]{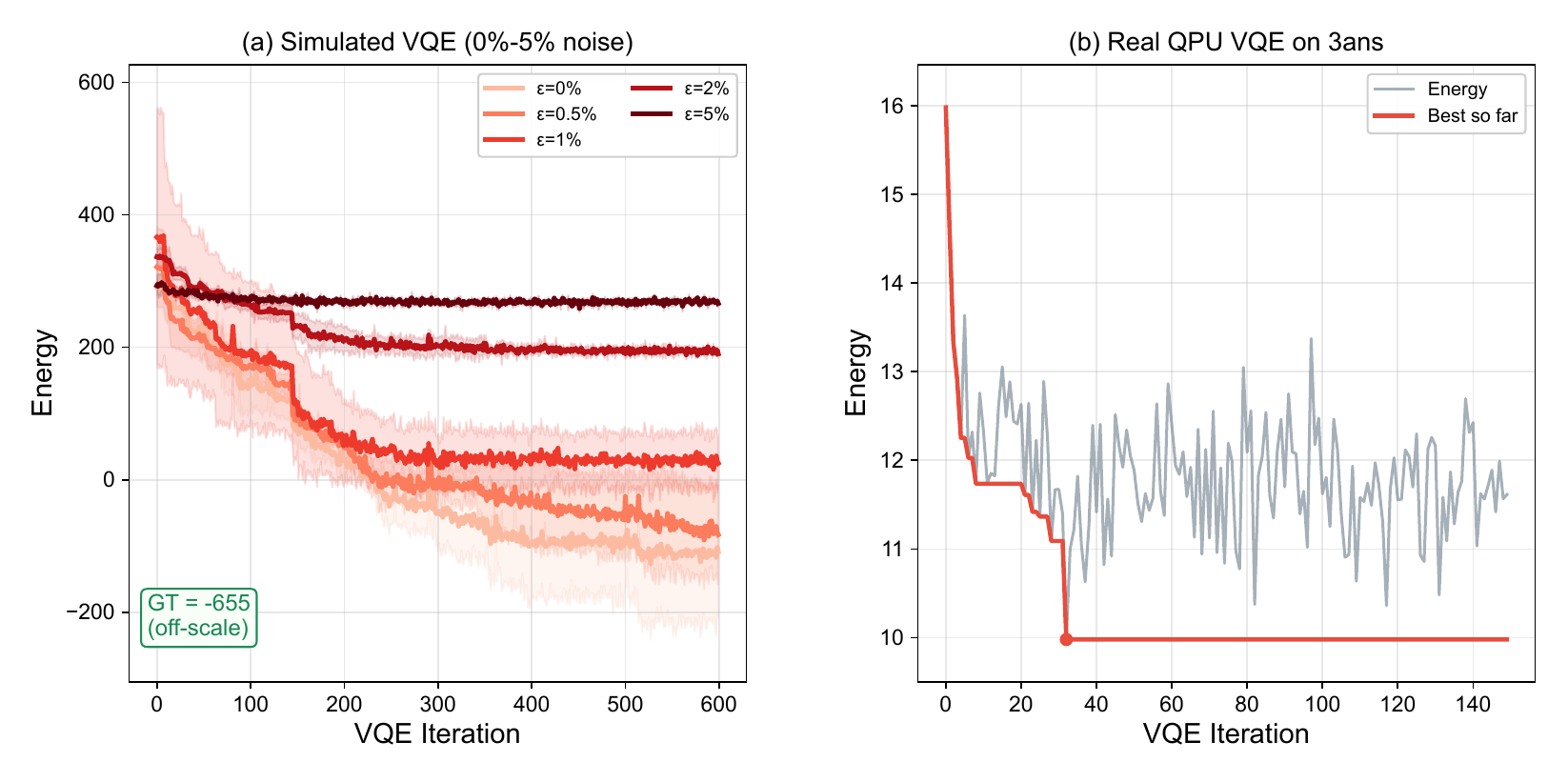}
\caption{VQE convergence trajectories. (a)~Simulated VQE on a 6-residue peptide under increasing per-gate depolarizing noise ($\varepsilon = 0\%$ to $5\%$). The ground-state energy ($-655$) lies far below the plotted range. At $\varepsilon \ge 2\%$, VQE cannot approach the ground state. (b)~Real QPU VQE on protein 3ans (IBM Heron R2). The measured energy (gray) oscillates throughout optimization; the best energy found (red) plateaus after ${\sim}25$ iterations with no further improvement.}
\label{fig:vqe_noise}
\end{figure}

Quantum computing offers an alternative route. The superposition property of a quantum processor allows it to explore all states in the eigenstate space mapped onto the measurement basis simultaneously~\cite{bauer2020quantum}. With an appropriate Hamiltonian formulation, quantum entanglement enables efficient exploration of the entire reachable conformational search space. However, current quantum methods still fall short. Variational quantum eigensolvers (VQE)~\cite{tang2021qubit}, which have been applied to protein structure prediction~\cite{zhang2026quantum,zhang2025hybrid} and can outperform deep learning in certain cases, are highly sensitive to quantum noise. As shown in Figure~\ref{fig:vqe_noise}(a), the convergence breaks down when the per-gate error exceeds 2\%, and on real hardware the optimizer exhibits persistent oscillations without stable improvement, as shown in Figure~\ref{fig:vqe_noise}(b). Non-iterative quantum methods such as sample-based quantum diagonalization (SQD)~\cite{robledo2025chemistry} and probabilistic quantum eigensolver (PQE)~\cite{seki2021quantum} avoid this optimization loop, but operate at the atomic orbital level and are not well aligned with the present problem. For an 18-residue peptide, they require 1,800--5,700 qubits, which is 12--39$\times$ beyond current hardware capabilities~\cite{robledo2025chemistry}.

We present \textit{QSAD} (\textbf{Q}uantum \textbf{S}ampling \textbf{A}nd \textbf{D}ecomposed Reconstruction), a quantum--classical framework designed to address this gap. QSAD formulates protein structure prediction through an amino-acid-level coarse-grained Hamiltonian, mapping the protein energy landscape into the eigenstate space while reducing the qubit cost to $O(N^2)$. Through multi-level one-way evolution and measurement, QSAD projects the energy landscape encoded in the eigenstate space onto a chosen measurement basis, represented as a distribution over candidate structures. Structural information is then inferred from ensemble statistics, enabling robustness to hardware noise without relying on error mitigation. QSAD is implemented as an end-to-end system that integrates quantum sampling with classical decoding, ranking, and reconstruction, producing both a predicted structure and an energy landscape projected onto the measurement basis.

We evaluate QSAD on 101 binding-pocket peptides of 5--18 residues with experimentally resolved crystal structures, executed entirely on the IBM Heron~R2 quantum processor. QSAD outperforms all evaluated baselines, improving prediction accuracy by 27--71\% over both AI and quantum methods while maintaining the lowest variance across all tested lengths. It also recovers the ground-state energy at noise levels 3--5$\times$ beyond typical device error rates, where VQE fails to converge, and reduces mean quantum execution time by $27\times$. These results show that amino-acid-level, non-iterative quantum sampling is a practical and effective computational path for short peptide structure prediction on current superconducting hardware.

Our contributions are as follows: \textbf{(1)} An amino-acid-level Hamiltonian encoding that reduces qubit requirements by modeling residue interactions through an implicit physical formulation, enabling mapping of the sequence conformational space into the quantum system; \textbf{(2)} A non-iterative quantum sampling architecture that replaces iterative optimization with one-way Hamiltonian evolution, achieving noise resilience through statistical patterns over measurement outcomes; \textbf{(3)} Experimental validation on 101 binding-pocket peptides executed on IBM Heron~R2, achieving a 27--71\% improvement over all evaluated AI and quantum baselines; \textbf{(4)} Reconstruction of an approximate protein energy landscape from the ensemble of measured structures, revealing the distribution of accessible conformational states encoded in the underlying eigenstate space.

\section{Background}

\subsection{Protein Folding and Energy Landscape}

Protein folding is governed by thermodynamics: a polypeptide chain adopts the conformation that minimizes its free energy~\cite{shakhnovich2006protein}. Predicting the native structure is therefore equivalent to finding the global minimum of an energy function that combines covalent geometry, non-covalent interactions, and solvent effects~\cite{plaxco2000topology}. The conformational space grows exponentially with chain length, making exhaustive search intractable~\cite{paquet2015molecular}. This energy function defines a high-dimensional surface over all possible conformations, known as the energy landscape~\cite{onuchic1997theory}. Its topology determines the native fold, folding kinetics, metastable states, and conformational flexibility. For binding-pocket peptides, the landscape is shaped by evolutionary pressure: sequences that cannot adopt pocket-compatible geometries do not persist as functional binding motifs~\cite{london2010structural}. The intrinsic energy landscape of such peptides therefore encodes binding-relevant structural information even without explicit modeling of the receptor. Characterizing this landscape as a conformational ensemble, rather than a single minimum, reveals the distribution of accessible states underlying binding affinity and mutational robustness~\cite{frauenfelder1991energy}.

\subsection{Quantum Computing and the Energy Landscape}

The protein folding energy landscape can be mapped onto a quantum system through a Hamiltonian~\cite{boulebnane2023peptide}. A Hamiltonian is essentially a function: it takes a candidate conformation as input, where that conformation is represented as a string of qubit values, and returns a single number, namely the energy of that conformation. Each possible fold of the peptide corresponds to a different qubit string, and the Hamiltonian scores it according to the underlying physics, such as whether neighboring residues form favorable contacts, whether any two residues collide, and whether the backbone geometry is physically plausible. The qubit string with the lowest score is called the ground state, and it corresponds to the most stable folded conformation. A quantum processor can prepare a superposition of many conformations simultaneously and measure it, thereby sampling the protein energy space mapped by this Hamiltonian. Sampling the Hamiltonian is therefore, in essence, equivalent to exploring the energy landscape.

\subsection{Evaluation Metric}

Structural prediction accuracy in protein science is measured by root-mean-square deviation (RMSD)~\cite{chothia1986relation}, reported in angstroms (\AA{}). RMSD is the sole universally accepted metric for comparing predicted and experimentally determined protein structures: it quantifies the average distance between corresponding atoms after optimal superposition. One angstrom equals 0.1~nm, roughly the length of a single covalent bond. Values below 3~\AA{} indicate high-accuracy predictions, 3--4~\AA{} reflects reasonable structural agreement, and values above 5~\AA{} generally indicate incorrect folds~\cite{dobson2003protein}. Throughout this paper, all reported RMSD values are computed over backbone C$\alpha$ atoms in the binding-pocket region. C$\alpha$ positions capture the overall backbone conformation, so computing C$\alpha$ distances effectively measures the global structural deviation of the predicted backbone.

\section{System Overview}
\label{sec:system_overview}

\begin{figure*}[htbp]
\centering
\includegraphics[width=0.9\textwidth]{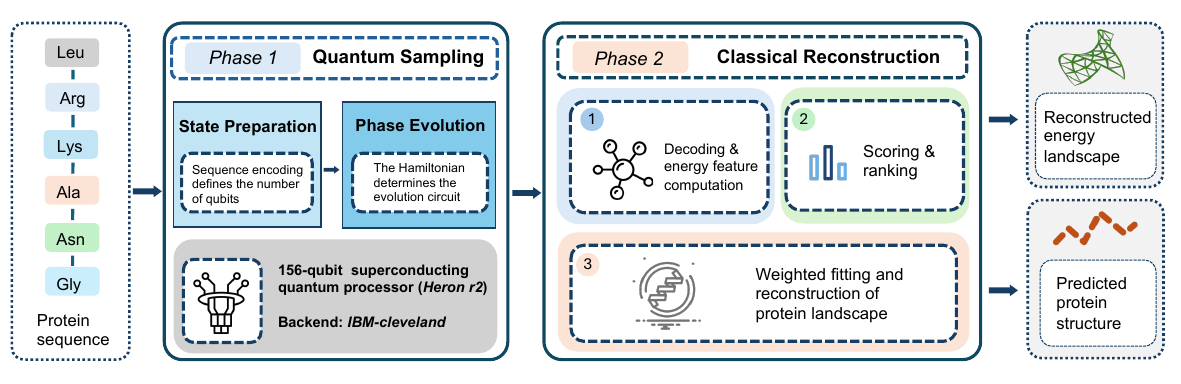}
\caption{End-to-end architecture of QSAD. \textbf{Left:} The input protein sequence is encoded into a coarse-grained Hamiltonian. \textbf{Center-left:} Quantum sampling executes on the IBM Heron~R2 processor (156 qubits), producing an ensemble of measured bitstrings. \textbf{Center-right:} Classical post-processing decodes bitstrings into lattice structures (Stage~1), scores and ranks candidates with diversity preservation (Stage~2), and reconstructs all-atom models (Stage~3). \textbf{Right:} The framework outputs both a predicted structure and a reconstructed energy landscape.}
\label{fig:system_overview}
\end{figure*}

QSAD takes an amino acid sequence as input and produces a predicted three-dimensional structure together with an approximate energy landscape. The system has two phases: quantum sampling on superconducting hardware, followed by classical post-processing that converts raw measurements into all-atom protein models. Figure~\ref{fig:system_overview} illustrates the architecture. Technical details of each component are given in Section~\ref{sec:system_design}.

\subsection{Phase 1: Quantum Sampling}
\label{sec:overview_quantum}
Given a peptide of length $N$, QSAD constructs a coarse-grained Hamiltonian that encodes folding physics into a qubit register. QSAD drives a single forward Hamiltonian evolution: randomized ansatz circuits prepare diverse initial states, and multiple evolution strengths $\beta$ are swept to sample different energy regimes. All circuits run independently with no classical feedback.

\subsection{Phase 2: Classical Reconstruction}
\label{sec:overview_classical}
The measured bitstrings are converted into all-atom structures through three stages: decoding into backbone C$\alpha$ coordinates (Stage~1), filtering, diversity-preserving selection, and physics-based ranking (Stage~2), and all-atom reconstruction (Stage~3). Backbone coordinates are fixed throughout this pipeline; reported RMSD values reflect the quantum-sampled lattice backbone directly. The sampled conformational ensemble is also fitted with a continuous energy surface, revealing energy basins, funnel structure, and the distribution of metastable states. By constructing a Hamiltonian based on the problem structure, the approximately reconstructed energy surface in the measurement basis can map, along a relevant dimension, all energy information accessible to the Hamiltonian. The resulting structures further support chemical validation and downstream applications.

\providecommand{\Id}{\mathbb{I}}
\providecommand{\Ind}{\Delta}

\section{System Design}
\label{sec:system_design}

\subsection{Coarse-Grained Protein Hamiltonian}
\label{sec:hamiltonian_design}

\subsubsection{Tetrahedral Lattice Representation}

QSAD models a peptide of length $N$ as a self-avoiding walk on a tetrahedral lattice. Each residue is a single backbone bead, and the folded conformation is determined by $N{-}1$ discrete turns, each taking one of four directions encoded by two binary variables:
\begin{equation}
  a \longmapsto \mathbf{v}(a) = \bigl(v^{(0)}(a),\, v^{(1)}(a)\bigr) \in \{0,1\}^2,
\end{equation}
so the configuration register requires $2(N{-}1)$ qubits. The tetrahedral lattice is chosen because its coordination number of four matches the $\mathrm{sp}^3$ backbone geometry, and each turn needs only two qubits~\cite{robert2021resource}.

\subsubsection{Implicit Side-Chain Model}

Side chains are not modeled as geometric variables. Their physicochemical effects are absorbed into sequence-dependent coefficients derived from a fixed amino acid feature table (volume, charge, hydrophobicity, aromaticity, hydrogen-bond tendency, conformational propensity). These features produce four coefficient classes: a \emph{pair coefficient} $\alpha_{ij}^{\mathrm{pair}}$ encoding contact preference via hydrophobic, electrostatic, and aromatic terms; a \emph{node coefficient} $\alpha_i^{\mathrm{node}}$ for burial/exposure preference; a \emph{local propensity} $\alpha_i^{\mathrm{loc}}$ for residue-specific backbone preferences (with special treatment of glycine and proline); and a \emph{steric coefficient} $\alpha_i^{\mathrm{ster}}$ for excluded volume. All weights are fixed constants derived from established residue contact potentials~\cite{miyazawa1993new, miyazawa1996residue}, lattice folding theory~\cite{dill1985theory}, and backbone conformational statistics~\cite{ramachandran1963stereochemistry}. No per-sequence tuning is needed; the same weight set is applied to all 101 proteins. This design eliminates explicit side-chain qubits and is the primary source of QSAD's qubit efficiency.

\subsubsection{Qubit Register Structure}

The Hilbert space factorizes as $\mathcal{H} = \mathcal{H}_{\mathrm{cont}} \otimes \mathcal{H}_{\mathrm{conf}}$, where $\mathcal{H}_{\mathrm{conf}}$ stores turn variables and $\mathcal{H}_{\mathrm{cont}}$ stores contact indicators. Admissible contacts are pairs $(i,j)$ with $j - i \ge 5$ and $j - i$ odd, reflecting the lattice bipartite structure. Each admissible pair gets a contact qubit $c_{ij}$ with projector $c_{ij} = (\Id - Z_{k_{ij}})/2$. After symmetry reduction (fixing the first two turns) and removing inactive qubits, the count ranges from 7 ($N{=}5$) to 145 ($N{=}18$).

\subsubsection{Hamiltonian Terms}

The complete Hamiltonian is:
\begin{equation}
  H = H_{\mathrm{back}} + H_{\mathrm{dist}} + H_{\mathrm{pair}} + H_{\mathrm{node}} + H_{\mathrm{loc}} + H_{\mathrm{ster}}.
  \label{eq:total_hamiltonian}
\end{equation}
$H_{\mathrm{back}} = \lambda_{\mathrm{back}} \sum_{i=1}^{N-2} T_{i,i+1}$ penalizes immediate backbone reversals, where $T_{pq}$ is the turn-alignment operator. $H_{\mathrm{dist}}$ enforces unit lattice distance at declared contacts via a primary penalty $\Phi_1(i,j) = \lambda_0(i,j)(D(i,j) - \Id)$ and second-neighbour overlap penalties. The four effective terms couple the registers through sequence-dependent coefficients: $H_{\mathrm{pair}} = \sum_{(i,j) \in \mathcal{A}} \alpha_{ij}^{\mathrm{pair}}\, \hat{c}_{ij}$ for contact preferences, $H_{\mathrm{node}} = \sum_i \alpha_i^{\mathrm{node}}\, \hat{N}_i$ for burial bias, and $H_{\mathrm{loc}}$, $H_{\mathrm{ster}}$ for local propensities and steric corrections.

\subsection{Quantum Sampling Circuit}
\label{sec:circuit_design}

\subsubsection{Circuit Architecture}

The sampling circuit applies two stages to the $n$-qubit register:
\begin{equation}
  |\Psi(s, \beta)\rangle = U_H(\beta)\; U_{\mathrm{ans}}(\boldsymbol{\theta}^{(s)})\; |0\rangle^{\otimes n},
  \label{eq:full_circuit}
\end{equation}
followed by computational-basis measurement.

\paragraph{State preparation}
The first stage uses an \texttt{EfficientSU2} ansatz with $R$ repetition layers and linear entanglement. Each layer $l$ applies single-qubit rotations and nearest-neighbour entangling gates:
\begin{equation}
  U_{\mathrm{ans}}^{(l)} = \prod_{j=1}^{n-1} \mathrm{CNOT}_{j,j+1} \;\cdot\; \bigotimes_{j=1}^{n} R_Z(\theta_{j,l}^{(z)})\, R_Y(\theta_{j,l}^{(y)}),
  \label{eq:ansatz_layer}
\end{equation}
so the full ansatz is $U_{\mathrm{ans}} = \prod_{l=1}^{R} U_{\mathrm{ans}}^{(l)}$ with $2nR$ parameters. The parameters $\boldsymbol{\theta}^{(s)}$ are \emph{not} optimized: for each seed $s$, they are drawn uniformly from $[0, 2\pi)$ using a deterministic random generator. Multiple seeds produce diverse initial states that explore different regions of the Hilbert space, trading optimization guarantees for sampling diversity.

\paragraph{Hamiltonian phase evolution}
The second stage applies
\begin{equation}
  U_H(\beta) = e^{-i\beta H_D}\, e^{-i\beta H}, \qquad H_D = \sum_{j=1}^{n} X_j,
\end{equation}
implemented via second-order Suzuki--Trotter decomposition~\cite{berry2007efficient} with $r$ repetitions. Multiple values of $\beta$ (default: $\{1, 2, 3, 4\}$) are swept to sample the landscape at different energy resolutions: small $\beta$ preserves broad exploration across conformational space, while large $\beta$ increases bias toward low-energy configurations.The physical mechanism is as follows. The phase kick $e^{-i\beta H}$ assigns each computational-basis component $|x\rangle$ a phase factor $e^{-i\beta E(x)}$, and the transverse driver $e^{-i\beta H_D}$ rotates the phase-encoded state out of the $Z$-eigenbasis so that these phase differences translate into $Z$-basis probability differences favouring low-energy conformations. As a result, the probability of observing a computational-basis bitstring $x$ depends on $\beta$ through
\begin{equation}
  p_x(\beta)=\left|\langle x|U_H(\beta)\,U_{\mathrm{ans}}|0\rangle^{\otimes n}\right|^2,
  \label{eq:beta_sampling}
\end{equation}
where $x$ denotes a measured bitstring. Different $\beta$ values therefore reshape the observed distribution in the measurement basis and emphasize different regions of the conformational landscape.A set of $\beta$ values spanning $[\beta_{\min}, \beta_{\max}]$ therefore covers a wider band of the energy landscape than any single value. In our experiments, we find that the results are not highly sensitive to the specific choice of $\beta$ values, provided that the set spans a sufficient range: the ablation study in Table~\ref{tab:sampling} confirms that the multi-$\beta$ strategy consistently outperforms any single $\beta$ in both coverage and effective sample size. The default set $\{1,2,3,4\}$ has not been optimized for individual proteins, and the same values are used across all 101 cases. Tuning $\beta$ per system or using a finer sweep could yield further improvements, which we leave to future work.

\subsubsection{Batch Execution and Transpilation}

\begin{figure}[htbp]
\centering
\includegraphics[width=1\linewidth]{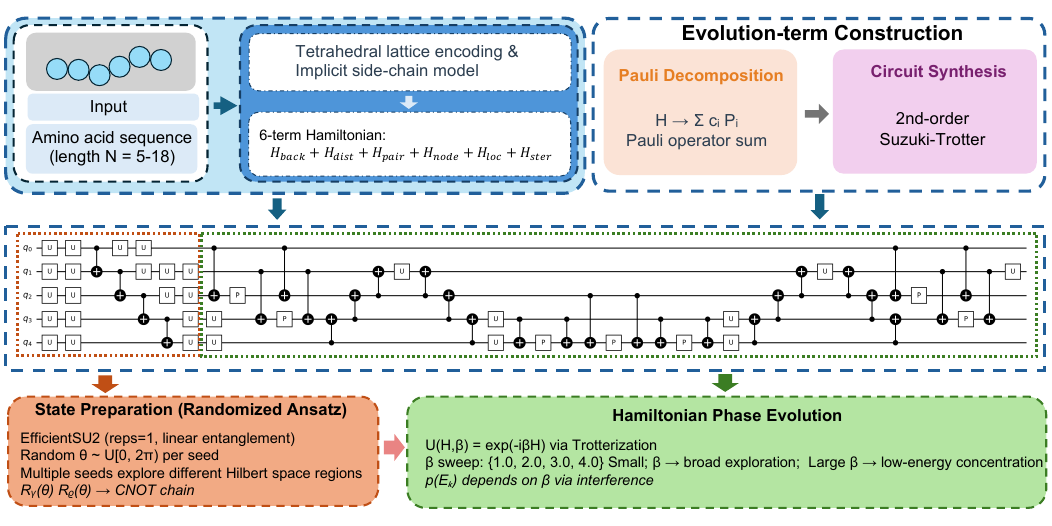}
\caption{Circuit-construction workflow of QSAD. The input amino-acid sequence is encoded to form a coarse-grained Hamiltonian. The Hamiltonian is decomposed into Pauli operators and compiled into a second-order Suzuki--Trotter circuit.}
\label{fig:sampling_stage}
\end{figure}

\begin{figure*}[htbp]
\centering
\includegraphics[width=\textwidth]{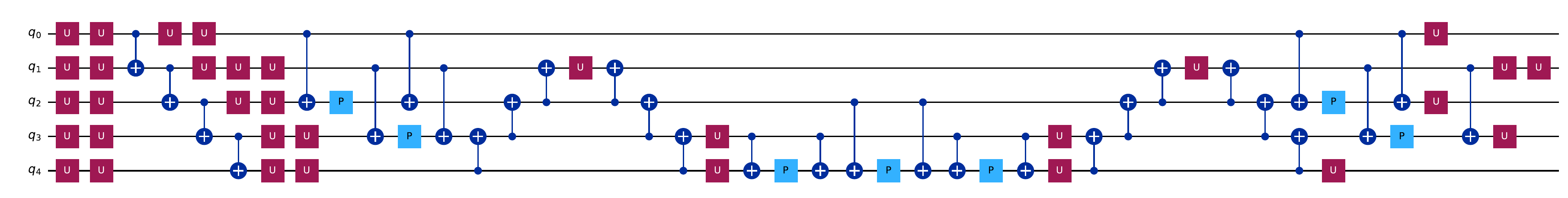}
\caption{Transpiled QSAD circuit for protein with 4 residues (5 qubits, $\beta = 2.0$) on IBM Heron~R2. Blue: single-qubit rotations; magenta: two-qubit ECR gates. Left portion: randomized ansatz; right portion: Trotterized Hamiltonian evolution.}
\label{fig:circuit}
\end{figure*}

For $S$ seeds and $|{\beta}|$ evolution strengths, QSAD builds sampling circuits by combining randomized state preparation with Hamiltonian phase evolution. The sequence is encoded as a coarse-grained Hamiltonian on the tetrahedral lattice, decomposed into Pauli operators, and compiled into a Trotterized circuit. Different seeds explore different regions of Hilbert space, while different $\beta$ values bias sampling toward low-energy states. With the default setting $S{=}3$, $\beta \in {1,2,3,4}$, and $G{=}10$, QSAD generates 120 circuits per protein. Each circuit uses 2,000 shots, for a total of 240,000 shots. Circuits in each group are submitted together as a batch job and transpiled to the IBM Heron~R2 native gate set at optimization level~3. Figure~\ref{fig:sampling_stage} shows the circuit-construction logic, and Figure~\ref{fig:circuit} shows an example transpiled circuit.

\subsection{Classical Post-Processing Pipeline}
\label{sec:pipeline_design}

\subsubsection{Step 1, Bitstring Decoding}

Each measured bitstring is decoded in three steps: (i)~the backbone configuration bits are extracted and mapped back to the turn sequence; (ii)~counts from bitstrings that differ only in contact-qubit assignments are aggregated to obtain the marginal frequency of each backbone conformation; (iii)~the turn sequence is converted into C$\alpha$ coordinates on the tetrahedral lattice. Coordinates are rescaled so that the median C$\alpha$--C$\alpha$ distance matches the physical peptide-bond length of 3.8~\AA{}. This aggregation step is important because the contact register is auxiliary: multiple contact-qubit patterns can correspond to the same backbone fold.

\subsubsection{Step 2, Filtering, Diversity Preservation, and Ranking}

Decoded conformations are first filtered for self-avoiding walk validity by rejecting any structure with non-adjacent bead overlaps. A greedy MaxMin procedure then selects diverse representatives: starting from the lowest-energy conformation, it iteratively adds the candidate most distant (in RMSD) from the current set, preserving distinct conformational basins that might otherwise be lost to energy-only ranking. Candidates are ranked by a composite physics-based score that sums six terms: backbone overlap penalty, Miyazawa--Jernigan contact energy~\cite{miyazawa1985estimation}, radius-of-gyration compactness, burial consistency (hydrophobic residues interior, polar residues exposed), contact order, and cross-group consensus (conformations found independently across multiple batch groups receive a bonus).

\subsubsection{Step 3, All-Atom Reconstruction}

Lattice C$\alpha$ coordinates are scaled to physical distances, and backbone heavy atoms (N, C$\alpha$, C, O) are placed using ideal peptide-bond geometry ($d_{\mathrm{C\alpha\text{-}C}} = 1.52$~\AA{}, $d_{\mathrm{C\text{-}N}} = 1.33$~\AA{}) via a NERF (Natural Extension Reference Frame)-style coordinate builder that positions each atom from the preceding three via bond length, bond angle, and dihedral angle~\cite{mackerell1998all}. Side chains are added from a backbone-dependent rotamer library. Backbone C$\alpha$ coordinates are fixed throughout this stage: reported RMSD values reflect the quantum-sampled lattice backbone directly, with no classical refinement bias. An optional Stage~4 supports full-atom energy minimization via OpenMM~\cite{eastman2023openmm} and PyRosetta~\cite{chaudhury2010pyrosetta} but is not used in this work.

\subsection{Energy Landscape Reconstruction}
\label{sec:landscape_design}
\begin{figure}[htbp]
\centering
\includegraphics[width=\linewidth]{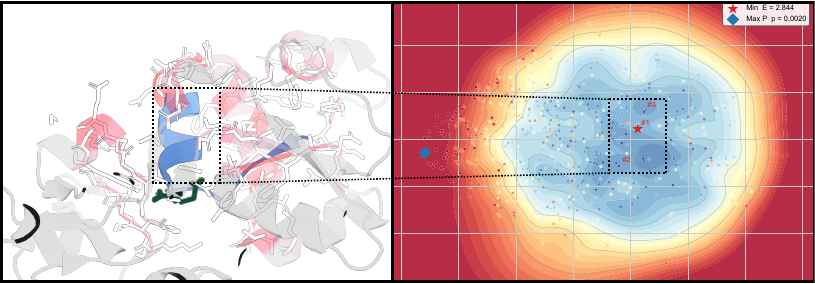}
\caption{Mapping the energy surface in protein structures using 2A5S (14 amino acids, pocket-region ligand-bound structure) as an example. The reconstructed energy space maps the contact potential of the real conformation by Euclidean distance, showing how the energy surface guides folding. In the left panel, red indicates high-energy regions and blue indicates low-energy regions. Because high-energy regions are already occupied by other atoms, the real conformation avoids them and instead lies in the blue low-energy region.}
\label{fig:mapping}
\end{figure}

Through the Hamiltonian, a native mapping can be established between the protein energy landscape and the eigenstate space. The energy modes in the eigenstate space correspond to the energy features on the protein energy surface. Finding the ground state in this eigenstate space with a quantum processor is equivalent to identifying the lowest-energy conformational mode on the protein energy landscape. At the same time, superposition allows these energy states to coexist simultaneously. Leveraging this property, the quantum system represents the entire protein energy surface defined by the Hamiltonian, and the resulting measurement distribution can be interpreted as a projection of this high-dimensional energy landscape onto the measurement basis. Figure~\ref{fig:mapping} illustrates the mapping between contact potential and conformational space. The lowest-energy conformations correspond to structures in which no steric conflicts occur in structural space.

Each valid conformation yields a coordinate vector $\mathbf{x}_k \in \mathbb{R}^{3N}$. PCA (Principal Component Analysis) on the set of $M$ valid conformations produces a 2-D embedding $\mathbf{z}_k$; in our experiments the top two components capture 60--75\% of structural variance. A continuous energy surface is fitted via thin-plate-spline RBF (Radial Basis Function interpolation) interpolation:
\begin{equation}
  \hat{E}(\mathbf{z}) = \sum_{k=1}^{M} w_k\, \phi\!\bigl(\|\mathbf{z} - \mathbf{z}_k\|\bigr),
\end{equation}
where $\phi(r) = r^2 \ln r$. Each landscape is characterized by a \emph{funnel score} (Pearson correlation between distance from the energy minimum and energy) and a \emph{basin count} (connected components on a gridded energy map at the 35th percentile). As a consistency check, we verify that sampling frequencies follow an approximate Boltzmann distribution~\cite{rowlinson2005maxwell} ($\ln p \approx -\beta_{\mathrm{eff}} E + \mathrm{const}$).

\section{Evaluation}
\label{sec:evaluation}

\subsection{Experimental Setup}
\label{sec:setup}

\paragraph{Dataset}
We evaluate QSAD on 101 binding-pocket peptide structures with experimentally resolved crystal structures, drawn from two sources: 55 cases from the QDockBank~\cite{zhang2025qdockbank} benchmark (5--14 residues, 7--97 qubits), which also have published VQE results for direct comparison, and 46 cases from PDBbind~\cite{wang2005pdbbind} (14--18 residues, 97--145 qubits). Lengths range from 5 to 18 residues with 3--18 cases per length.

\paragraph{Hardware and baselines}
All QSAD experiments run on the IBM Heron~R2 processor (\texttt{ibm\_cleveland}, 156 qubits), using 120 circuits at 2{,}000 shots each (240{,}000 total shots per protein) submitted as 10 independent batch jobs.

We compare against seven baselines spanning three paradigms. \textit{(i)~Multiple Sequence Alignment(MSA)-based:} ColabFold-MSA~\cite{mirdita2022colabfold} uses the AlphaFold2 architecture with a full MSA pipeline (MMseqs2, ${\sim}500$\,GB database), inferring residue contacts from co-evolutionary patterns across homologous sequences. \textit{(ii)~Single-sequence:} ESMFold~\cite{lin2022language,rives2019biological} and OmegaFold~\cite{OmegaFold} extract structural information from protein language models; ColabFold~\cite{kim2025easy} and OpenFold~\cite{Ahdritz2022} use the AlphaFold2 architecture in single-sequence mode to test whether the structure module alone suffices for short peptides. \textit{(iii)~Diffusion-based:} AlphaFold3~\cite{abramson2024accurate} replaces AF2's structure module with a diffusion model that iteratively denoises atomic coordinates, representing the current state of the art. On the quantum side, we compare against VQE from the QDockBank benchmark (55 cases, up to 14 residues).

\paragraph{Metric}
Structural quality is measured by backbone C$\alpha$ RMSD~\cite{chothia1986relation} against the crystal structure, computed over the binding-pocket region only.

\subsection{QPU Execution Summary}
\label{sec:runtime}

\begin{table}[htbp]
\centering
\caption{Ansatz depth and mean QPU time by peptide length. Ansatz uses EfficientSU2 ($\mathrm{reps}{=}1$, linear entanglement). Values in parentheses denote the number of cases.}
\label{tab:runtime}
\begingroup
\small
\setlength{\tabcolsep}{4pt}
\renewcommand{\arraystretch}{1.08}
\begin{adjustbox}{max width=\linewidth}
\begin{tabular}{rrrr|rrrr}
\toprule
$N$ & Depth (cases) & Qubits & QPU time & $N$ & Depth (cases) & Qubits & QPU time \\
\midrule
5  & 10(4)   &   7 &  0.5 & 12 &  80(3)  &  77 &  9.0 \\
6  & 21(6)   &  18 &  0.9 & 13 &  90(3)  &  87 & 12.6 \\
7  & 36(3)   &  33 &  1.8 & 14 & 100(18) &  97 & 16.0 \\
8  & 44(8)   &  41 &  2.4 & 15 & 111(9)  & 108 & 20.4 \\
9  & 52(4)   &  49 &  3.6 & 16 & 123(9)  & 120 & 26.3 \\
10 & 61(11)  &  58 &  4.8 & 17 & 135(9)  & 132 & 33.4 \\
11 & 70(5)   &  67 &  6.3 & 18 & 148(9)  & 145 & 44.1 \\
\bottomrule
\end{tabular}
\end{adjustbox}
\endgroup
\end{table}

The 120 circuits per protein (Section~\ref{sec:setup}) produce 24.2~million shots across the full campaign. Table~\ref{tab:runtime} lists the ansatz circuit depth for each peptide length. The \texttt{EfficientSU2} ansatz ($\mathrm{reps}{=}1$, linear entanglement) grows from depth 10 (5 residues, 7 qubits) to 148 (18 residues, 145 qubits). For the 101 proteins with complete IBM Runtime telemetry, the total quantum execution time is 23.0~hours (mean 22.3~min per protein), with a mean queue wait of 3.3~s per job group, indicating near-zero queuing overhead under batch submission mode. The total classical post-processing time is 1.1~hours across all 101 proteins (mean 5.9~min per protein).

\subsection{Structural Prediction Accuracy}
\label{sec:accuracy}

For each of the 101 proteins, QSAD runs the full pipeline: quantum sampling on Heron~R2, followed by classical post-processing (decoding, filtering, scoring, and all-atom reconstruction). Backbone coordinates are fixed during reconstruction, and the optional refinement stage is not used. All reported RMSD values therefore reflect the quantum-sampled lattice backbone directly. AI baselines are evaluated with default configurations, and VQE results are taken from the published QDockBank benchmark.

\begin{figure*}[htbp]
\centering
\includegraphics[width=0.95\textwidth]{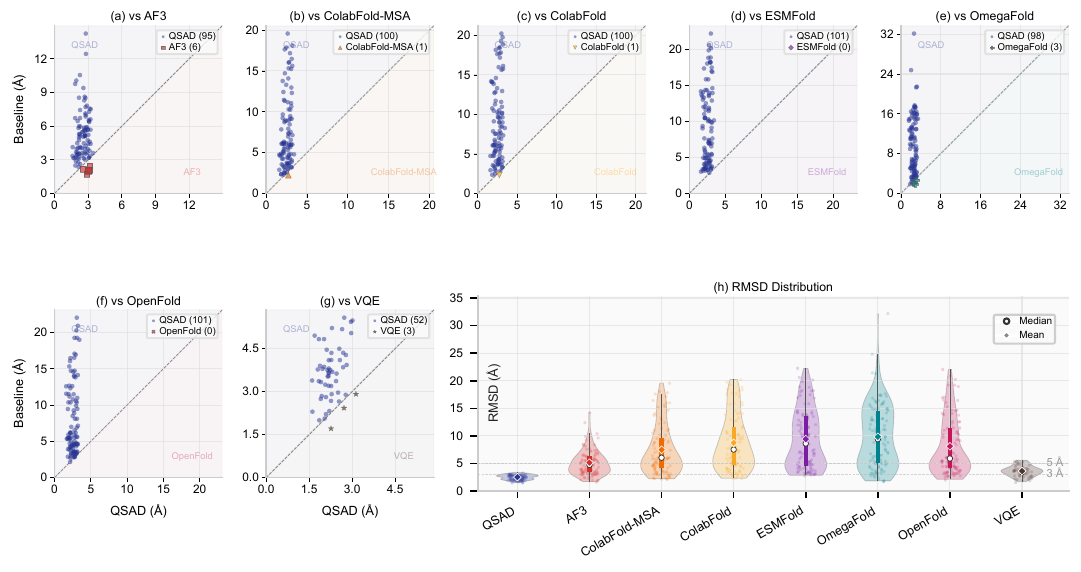}
\caption{Comparison of QSAD against structure-prediction baselines across 101 proteins. (a)--(g) Pairwise RMSD comparisons between QSAD and each baseline; points above the diagonal indicate cases where QSAD achieves lower RMSD. The legends report win counts for each method pair: 95/101 vs.\ AF3, 100/101 vs.\ ColabFold-MSA, 100/101 vs.\ ColabFold, 101/101 vs.\ ESMFold, 98/101 vs.\ OmegaFold, 101/101 vs.\ OpenFold, and 52/55 vs.\ VQE on shared cases. (h) RMSD distributions for all methods, with violin plots showing the full spread and markers indicating mean and median values. QSAD has the lowest central error and the narrowest distribution among all compared methods.}
\label{fig:scatter}
\end{figure*}

Figure~\ref{fig:scatter} summarizes both pairwise and distributional comparisons against all baselines. In the scatter panels, most points lie above the diagonal, showing that QSAD achieves lower RMSD than AF3, ColabFold-MSA, ColabFold, ESMFold, OmegaFold, OpenFold, and VQE on the large majority of cases. The violin panel further shows that QSAD has the lowest median RMSD at 2.7~\AA{} and mean RMSD at 2.6~\AA{}, compared with AF3 at 4.8~\AA{} (mean 5.2~\AA{}), ColabFold-MSA at 6.1~\AA{}, ColabFold at 7.5~\AA{}, ESMFold at 8.6~\AA{}, OmegaFold at 9.4~\AA{}, and OpenFold at 5.9~\AA{}. QSAD also exhibits the tightest distribution, with RMSD concentrated in a narrow low-error range, whereas all other baselines show substantially broader spreads. VQE reaches a median RMSD of 3.7~\AA{} on its 55 shared cases, but remains limited to shorter proteins and shows larger variability.

\subsection{Execution Time and Scalability}
\label{sec:time}

We compare execution time on the 55 shared QDockBank cases. QSAD completes each case with a mean of 1{,}966~s (33~min) and median 1{,}470~s (25~min). VQE requires a mean of 28{,}036~s (467~min) and median 9{,}870~s (164~min). QSAD is faster on all 55 cases, with a mean speedup of $27\times$ and a maximum of $365\times$.

\begin{figure}[htbp]
\centering
\includegraphics[width=\columnwidth]{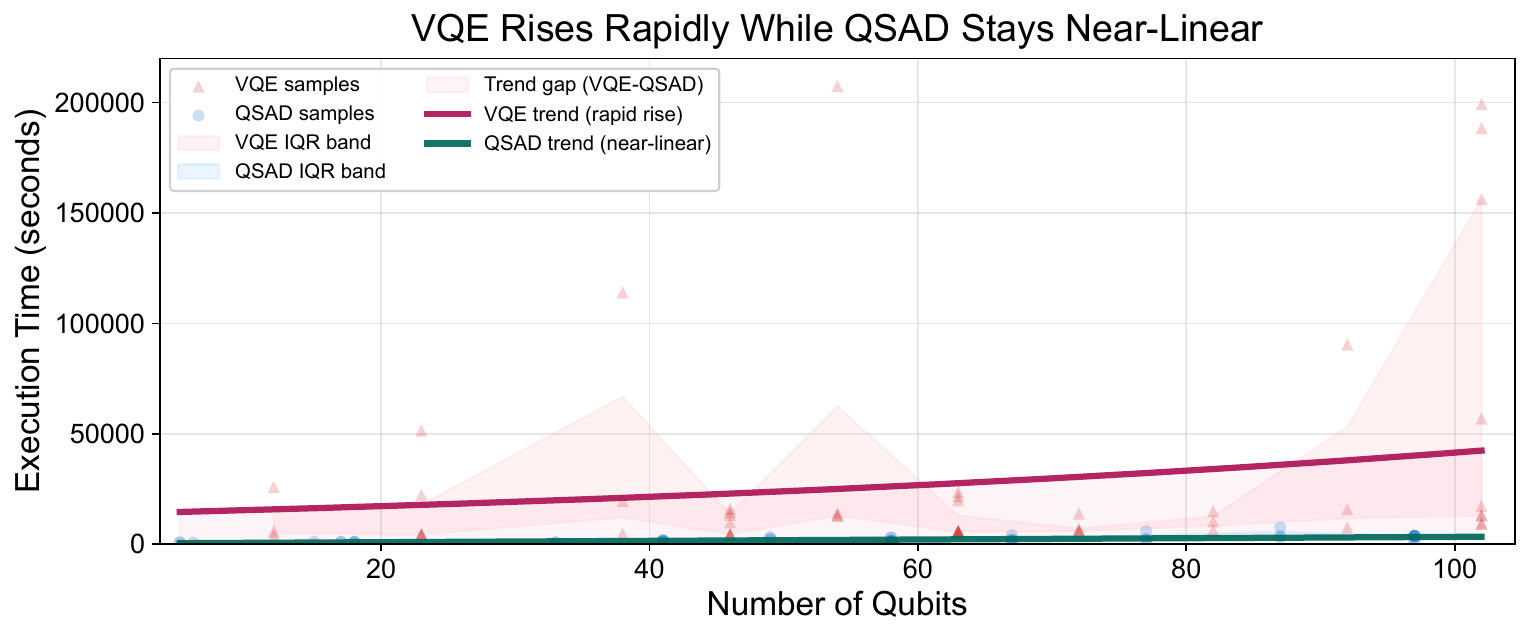}
\caption{Execution time vs.\ qubit count. QSAD shows linear growth with bounded runtime (max 2.1~h); VQE shows unpredictable runtime (max 57.6~h).}
\label{fig:timing}
\end{figure}

QSAD's runtime scales approximately linearly with qubit count, with low variance across groups (CV (Coefficient of Variation) $\approx 0.12$). VQE shows unpredictable scaling: individual cases range from 67~min to over 57~hours, because convergence depends on noise, optimizer trajectory, and landscape ruggedness (Figure~\ref{fig:timing}). In practical deployment, unpredictable runtime is a significant obstacle for task scheduling and resource planning. The difference in runtime stability between VQE and QSAD stems from their execution models: VQE relies on a session mode that maintains a long-lived bidirectional connection to the QPU, while QSAD submits independent batch jobs with no return communication, making its runtime predictable and controllable.

\subsection{Energy Landscape Analysis}
\label{sec:landscape_eval}

We reconstruct energy landscapes for all 101 proteins (5--18 residues) with at least 40 valid conformations each (mean 947 per protein). The sampled ensemble is projected via PCA and fitted with a continuous energy surface.

\begin{figure*}[htbp]
\centering
\includegraphics[width=0.9\textwidth]{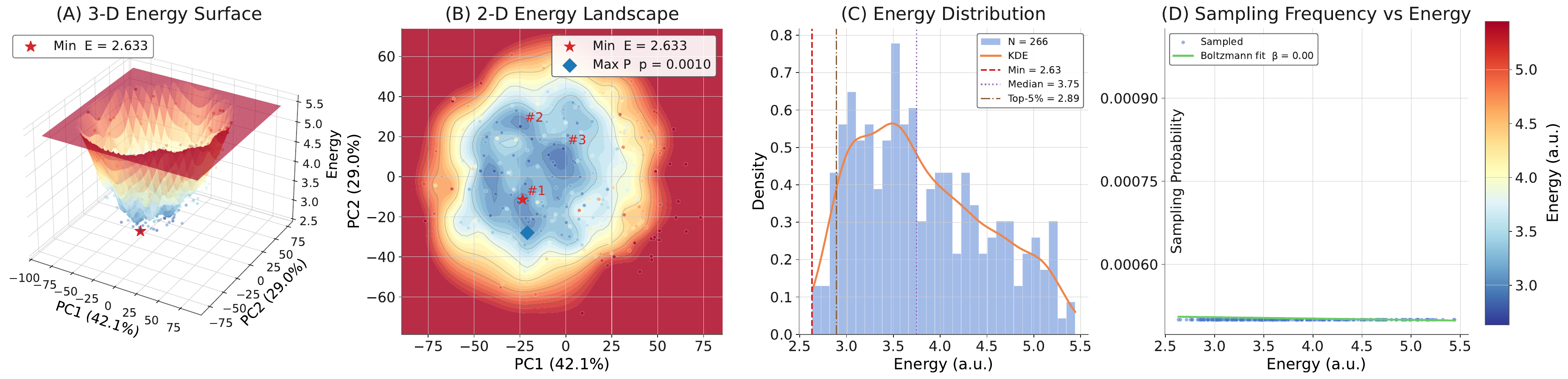}
\caption{Reconstructed energy landscape for protein 1A0Q (16 residues, 120 qubits, funnel score 0.745). (A) 3-D fitted energy surface. (B) 2-D contour in PCA space; the red star marks the global minimum and the blue diamond marks the maximum sampling-probability point. (C) Energy distribution of sampled conformations. (D) Sampling frequency versus energy with a fitted Boltzmann trend ($\beta_{\mathrm{eff}}=0.005$), consistent with a dominant funnel.}
\label{fig:landscape_1a0q}
\end{figure*}

\begin{figure*}[htbp]
\centering
\includegraphics[width=0.9\textwidth]{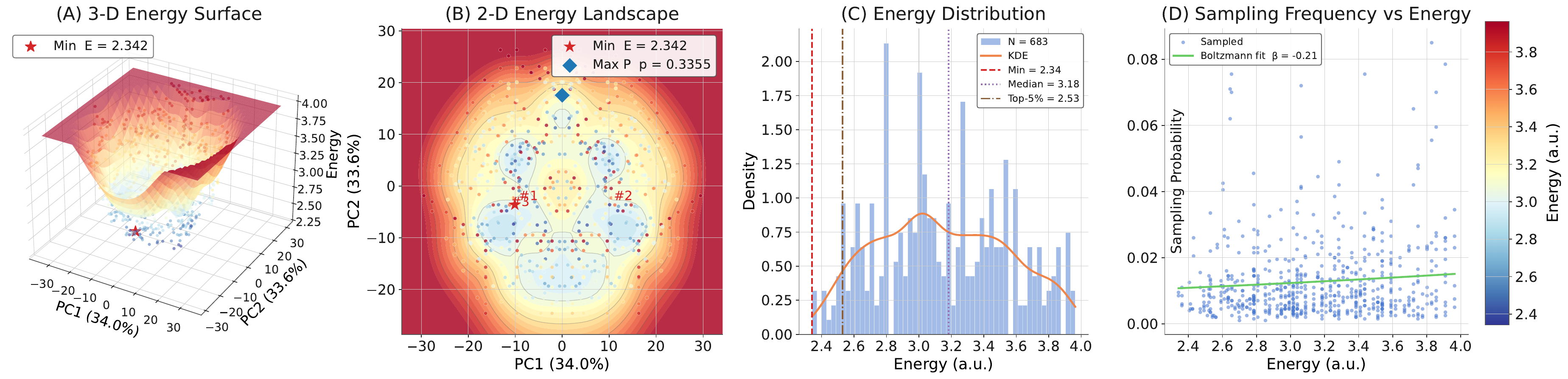}
\caption{Reconstructed energy landscape for protein 4Y79 (9 residues, 49 qubits, funnel score 0.261). (A) 3-D fitted energy surface. (B) 2-D contour in PCA space; multiple local basins are visible, with the maximum sampling-probability point separated from the global minimum. (C) Energy distribution of sampled conformations. (D) Sampling frequency versus energy with a fitted Boltzmann trend ($\beta_{\mathrm{eff}}=-0.209$), indicating a broader multi-basin landscape.}
\label{fig:landscape_4y79}
\end{figure*}

Across the 101 landscapes, the mean funnel score, the definition in section~\ref{sec:landscape_design}, is 0.619 (median 0.698), with basin counts ranging from 1 to 21 (mean 6.1).Different basin counts correspond to distinct energy landscape topologies and distinct chemical preferences; complex landscapes further explain why most methods fail to produce accurate structures. Figure~\ref{fig:landscape_1a0q} shows protein 1A0Q (16 residues, funnel score 0.745): a clear primary funnel with subsidiary basins, and a Boltzmann-consistent sampling distribution ($\beta_{\mathrm{eff}} = 0.005$). Figure~\ref{fig:landscape_4y79} shows a contrasting case, protein 4Y79 (9 residues, funnel score 0.261): a rough, multi-basin landscape suggesting multiple competing conformations. These two cases illustrate that the landscape reconstruction captures qualitatively different conformational regimes. 
The reconstructed landscapes carry practical value beyond structural prediction. A well-funnelled landscape (high funnel score) suggests that the peptide adopts a dominant bound-state conformation, while a rough, multi-basin landscape may indicate conformational heterogeneity relevant to binding promiscuity or allosteric regulation. Such information is inaccessible to single-structure predictors, which return one conformation and provide no view of the surrounding energy surface. The ability to distinguish funnelled from frustrated landscapes from a single round of quantum sampling is a unique capability of the QSAD framework.

\subsection{Noise Resilience}
\label{sec:noise}

\paragraph{Fairness and Reproducibility Statement}
In this set of experiments, running on real QPUs would affect both reproducibility and fairness, because noise on physical quantum backends fluctuates over time and with environmental conditions, and noise levels also vary across physical qubits. In addition, to study system behavior under different noise levels, we cannot directly tune the noise on a real backend, since its noise profile remains within a relatively stable and narrow range. Therefore, under identical conditions, we conduct fair and reproducible experiments on a simulator by emulating different noise levels for both methods.

\paragraph{Setup}

\begin{table}[htbp]
\centering
\caption{Noise levels and real-hardware equivalences. Signal retention $\alpha = (1-\varepsilon)^d$ for VQE ($d \approx 29$) and QSAD ($d \approx 30$--$55$).}
\label{tab:noise_config}
\begingroup
\small
\setlength{\tabcolsep}{4pt}
\renewcommand{\arraystretch}{1.08}
\begin{adjustbox}{max width=\linewidth}
\begin{tabular}{ccccl}
\toprule
$\varepsilon$ & Label & VQE $\alpha$ & QSAD $\alpha$ & Real hardware comparison \\
\midrule
0\%   & Fault-tolerant & 1.000 & 1.000       & Noiseless simulator \\
0.5\% & Near-ideal     & 0.865 & 0.76--0.86  & $\approx$ Best trapped-ion 2Q gates \\
1\%   & Low            & 0.747 & 0.58--0.74  & $\approx$ IBM 1Q gate; ion-trap 2Q gate \\
2\%   & NISQ           & 0.557 & 0.33--0.55  & $\approx$ IBM 2Q gate (typical NISQ) \\
\bottomrule
\end{tabular}
\end{adjustbox}
\endgroup
\end{table}
We compare QSAD and VQE under progressively increasing per-gate depolarizing noise on a 6-residue test system (SAASAS, 18 qubits). Each gate depolarizes the state with probability $\varepsilon$, so a circuit of depth $d$ retains signal fraction $\alpha = (1-\varepsilon)^d$. We test four levels from $\varepsilon = 0\%$ (ideal) to $\varepsilon = 2\%$, spanning noiseless, near-ideal, low-noise, and typical NISQ conditions. Table~\ref{tab:noise_config} maps each level to its real-hardware equivalence and signal retention. VQE uses the same \texttt{EfficientSU2} ansatz with COBYLA (600 iterations, 4{,}096 shots per iteration, ${\sim}2.5 \times 10^6$ total shots, 3 repeats per level). QSAD uses the stratified ensemble (48 circuits, 49{,}152 total shots).

\paragraph{Results}

\begin{figure}[htbp]
\centering
\includegraphics[width=\columnwidth]{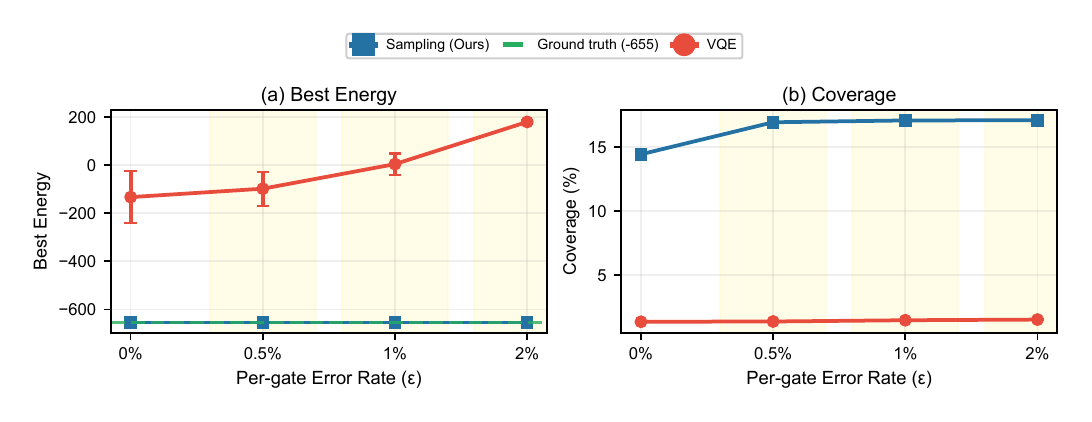}
\caption{Noise resilience comparison. (a) Best energy vs.\ error rate: QSAD consistently recovers the ground truth, while VQE degrades above $\varepsilon = 1\%$. (b) Coverage: QSAD maintains $>$14\% at all levels, while VQE stays below 2\%.}
\label{fig:noise_resilience}
\end{figure}


Table~\ref{tab:noise_comparison} presents the detailed comparison. The two methods respond to noise in opposite ways. VQE's low-energy coverage drops from 2.9\% to 1.3\%, and energy estimates show increasing variance above $\varepsilon = 1\%$. At $\varepsilon = 2\%$, the typical NISQ regime, VQE convergence becomes unstable. Under ideal conditions, COBYLA converges within approximately 200 iterations, but at $\varepsilon = 2\%$ the optimizer oscillates without settling (Figure~\ref{fig:vqe_noise}).

\begin{table}[htbp]
\centering
\caption{QSAD and VQE performance under noise. VQE columns show mean $\pm$ std over 3 repeats.}
\label{tab:noise_comparison}
\begingroup
\small
\setlength{\tabcolsep}{4pt}
\renewcommand{\arraystretch}{1.08}
\begin{adjustbox}{max width=\linewidth}
\begin{tabular}{c rrrr r@{$\;\pm\;$}lr@{$\;\pm\;$}l}
\toprule
& \multicolumn{4}{c}{\textbf{QSAD}} & \multicolumn{4}{c}{\textbf{VQE}} \\
\cmidrule(lr){2-5} \cmidrule(lr){6-9}
$\varepsilon$ & Unique & Cov. & Low-E & Best $E$ & \multicolumn{2}{c}{Best $E$} & \multicolumn{2}{c}{Low-E (\%)} \\
\midrule
0\%   & 37{,}892 & 14.5\% & 13.1\% & $-655.0$ & $-655.0$ & $0.0$ & $2.91$ & $0.85$ \\
0.5\% & 44{,}459 & 17.0\% & 16.8\% & $-655.0$ & $-655.0$ & $0.0$ & $2.42$ & $0.97$ \\
1\%   & 44{,}834 & 17.1\% & 17.3\% & $-655.0$ & $-653.8$ & $1.0$ & $2.13$ & $0.34$ \\
2\%   & 44{,}872 & 17.1\% & 17.3\% & $-655.0$ & $-654.2$ & $1.2$ & $1.27$ & $0.08$ \\
\bottomrule
\end{tabular}
\end{adjustbox}
\endgroup
\end{table}

QSAD shows the opposite behavior: coverage \emph{increases} from 14.5\% to 17.1\% under noise, because depolarizing noise acts as a perturbation that broadens conformational exploration. QSAD recovers the ground-state energy ($E^* = -655.0$) at every noise level tested up to $\varepsilon = 2\%$. This robustness follows from the sampling architecture: corrupted individual measurements do not compound, because the ensemble averages them out instead of feeding them back into an optimization loop.


\subsection{System Design Validation}
\label{sec:system_validation}

The preceding sections establish that QSAD achieves high accuracy, fast execution, landscape reconstruction, and noise resilience. This section examines three questions about the origin of these results: where in the pipeline does the structural advantage arise, how does the sampling strategy contribute, and does quantum sampling outperform classical alternatives on the same Hamiltonian.

\begin{figure}[htbp]
\centering
\includegraphics[width=\columnwidth]{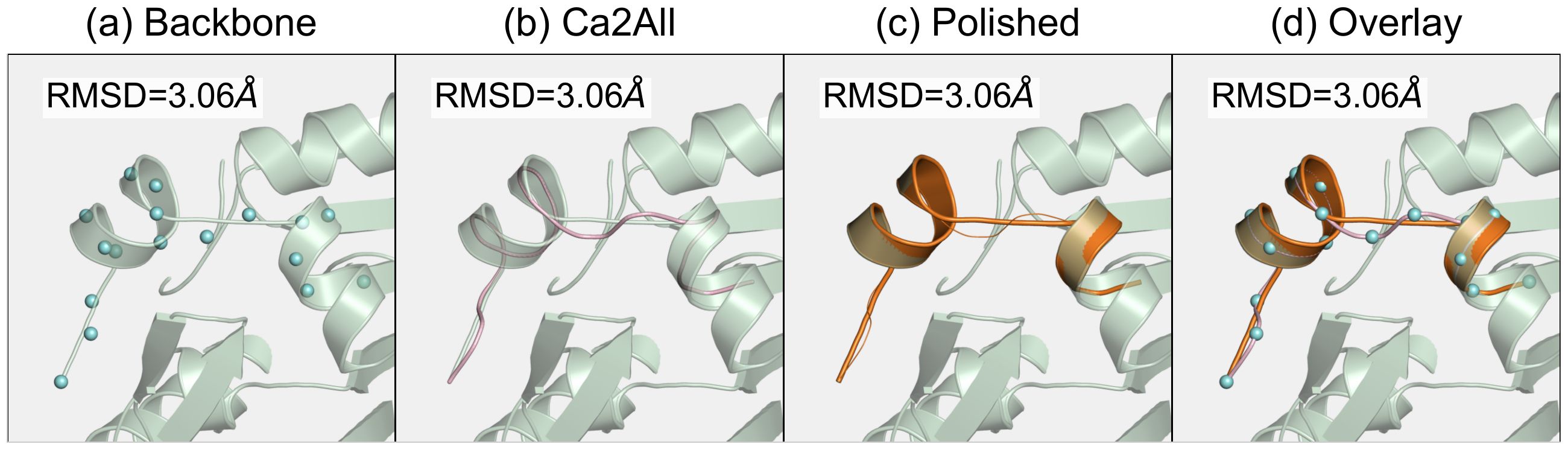}
\caption{Structure overlay for protein 2A14 across three pipeline stages. The C$\alpha$ backbone remains identical throughout. Panels a--d show the progression from backbone nodes to full-atom reconstruction and their overlap. No structural change or displacement is introduced during post-processing, and post-processing itself does not improve the protein structure.}
\label{fig:exp2a}
\end{figure}

\begin{figure}[htbp]
\centering
\includegraphics[width=\columnwidth]{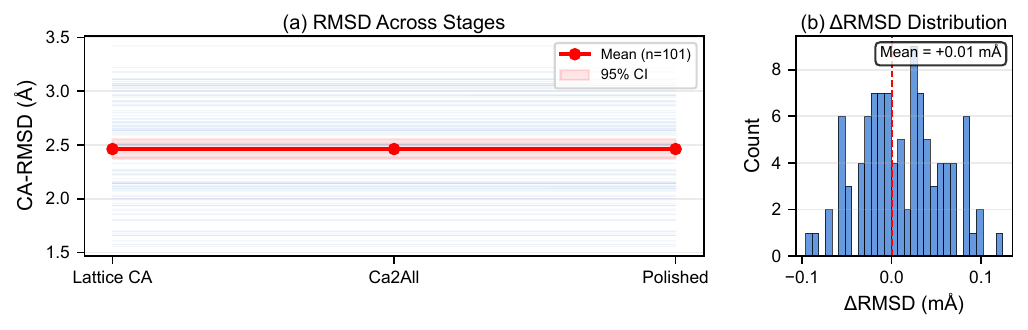}
\caption{Per-protein RMSD across three pipeline stages (101 proteins). The band is flat: post-processing introduces negligible change ($< 0.0002$~\AA{}).}
\label{fig:exp2b}
\end{figure}

\paragraph{Advantage origin}
We track RMSD through three pipeline stages: (1)~lattice C$\alpha$ trace from coarse scoring, (2)~after all-atom reconstruction, and (3)~after optional energy minimization. Across all 101 proteins, the mean RMSD is 2.462~\AA{} at all three stages. The all-atom step adds backbone heavy atoms and side chains around the existing C$\alpha$ trace but does not move C$\alpha$ positions. Energy minimization shifts coordinates by at most 0.0001~\AA{}. Figure~\ref{fig:exp2a} illustrates this for protein 2A14; Figure~\ref{fig:exp2b} confirms the flat RMSD band across all 101 cases. The structural quality is determined entirely at the quantum sampling stage; post-processing contributes atomic detail without altering the backbone geometry.

\paragraph{Sampling strategy ablation}
We validate that each component of QSAD's sampling strategy contributes to conformational exploration through a controlled ablation on a 6-residue peptide (SAASAS, 18 qubits, $2^{18} = 262{,}144$ conformations). Three configurations are compared under the same shot budget (49{,}152 shots): \textbf{L0 (Random)}: Ansatz only, $\beta = 0$, single seed. \textbf{L1 (Single Evolved)}: Single seed, $\beta = 1.0$. \textbf{L2 (Stratified)}: 3~groups $\times$ 4~seeds $\times$ 4~$\beta$ values, the full QSAD strategy.

\begin{figure}[htbp]
\centering
\includegraphics[width=\columnwidth]{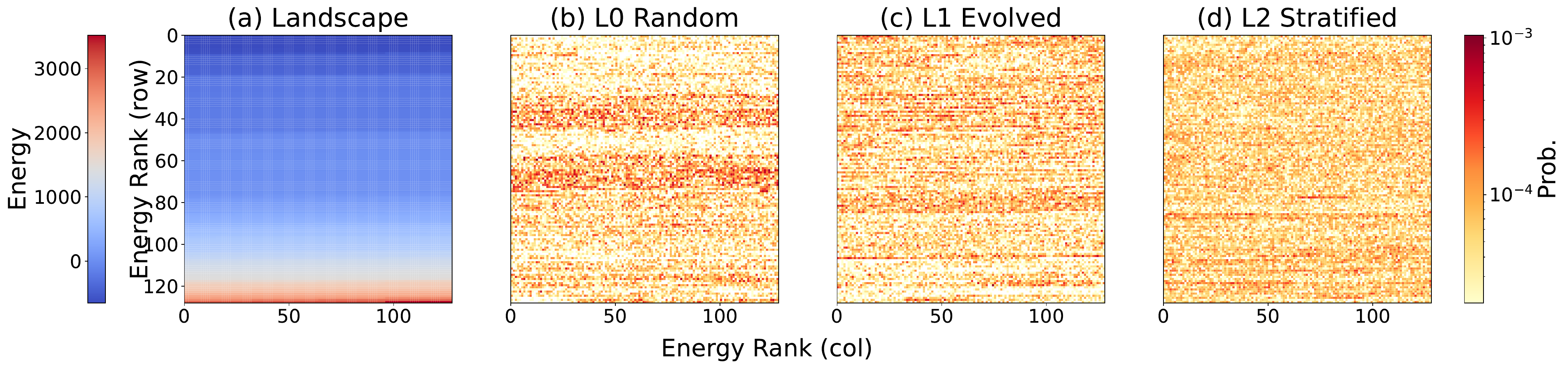}
\caption{Sampling heatmaps for three strategy levels versus the true energy landscape. Panel (a) shows the ground-truth landscape ordered by energy rank. Panels (b)–(d) show the sampled probability maps for L0 random, L1 single-seed evolved, and L2 stratified sampling under the same shot budget. From L0 to L2, sampling becomes less concentrated and covers a broader energy range, indicating improved conformational exploration. L2 provides the most balanced coverage across low- and mid-energy regions.}
\label{fig:heatmap}
\end{figure}

\begin{figure}[htbp]
\centering
\includegraphics[width=0.8\columnwidth]{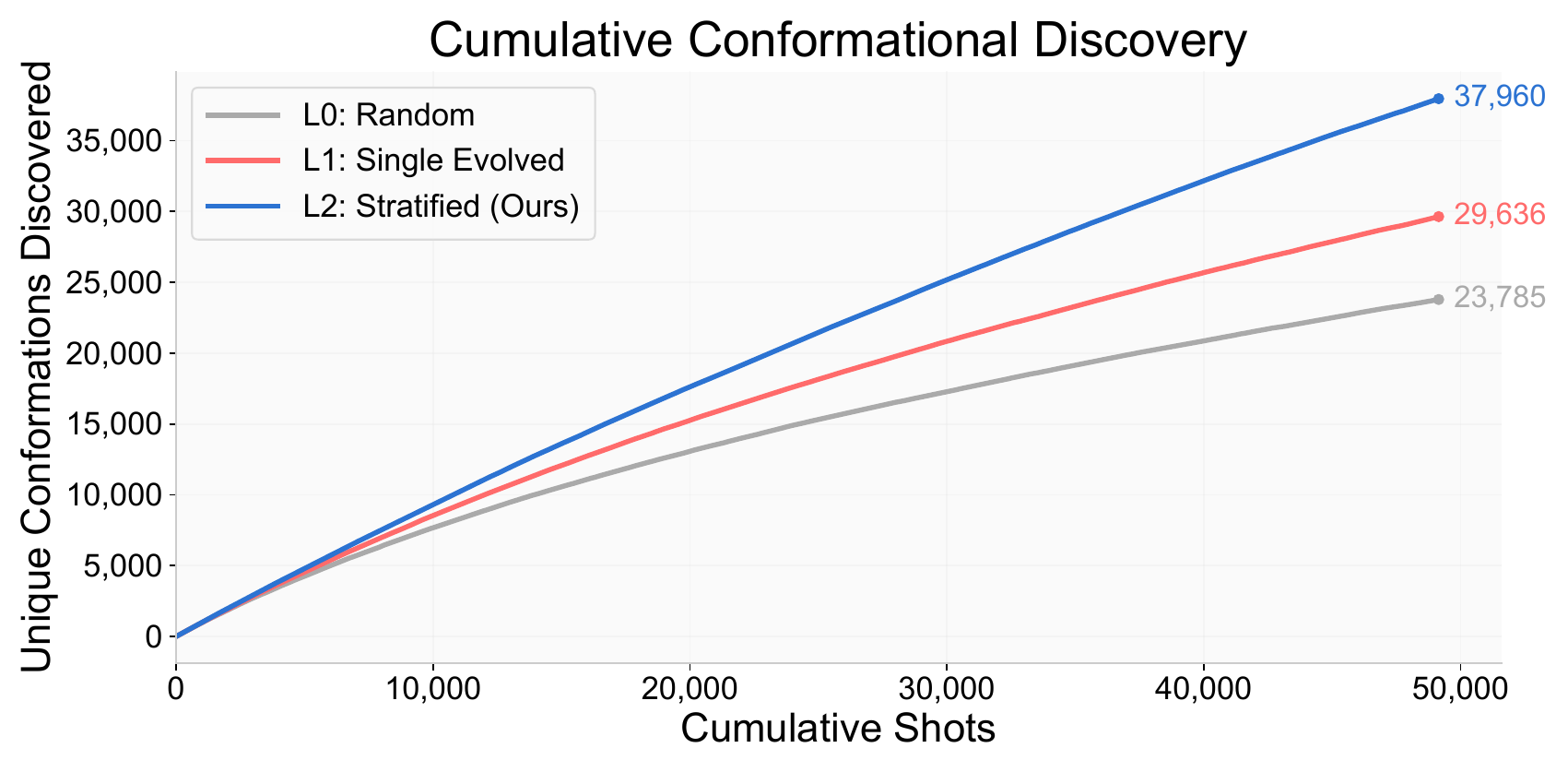}
\caption{Cumulative unique conformations discovered vs.\ total shots. L2 maintains a consistently higher discovery rate.}
\label{fig:cumulative}
\end{figure}

Table~\ref{tab:sampling} shows the results. All three configurations find the ground-truth energy ($E^* = -655.0$), but coverage quality differs substantially. Adding Hamiltonian evolution (L0 $\to$ L1) increases low-energy coverage from 5.5\% to 12.6\% and reduces mean energy from 272.7 to 179.2, confirming that evolution biases sampling toward low-energy regions. Stratification (L1 $\to$ L2) further raises total coverage to 14.5\% and ESS (Effective Sample Size) by $1.64\times$, by exploring multiple energy regimes simultaneously. Figure~\ref{fig:heatmap} visualizes the progressive improvement; Figure~\ref{fig:cumulative} shows L2's superior discovery rate.

\begin{table}[htbp]
\centering
\caption{Sampling strategy comparison (49{,}152 shots each). All methods found $E^* = -655.0$. (Higher is better)}
\label{tab:sampling}
\begingroup
\small
\setlength{\tabcolsep}{4pt}
\renewcommand{\arraystretch}{1.08}
\begin{adjustbox}{max width=\linewidth}
\begin{tabular}{lrrrrrr}
\toprule
\textbf{Method} & \textbf{Unique} & \textbf{Coverage} & \textbf{Low-E} & \textbf{Mean $E$} & \textbf{Entropy} & \textbf{ESS} \\
\midrule
L0: Random      & 23{,}785 &  9.1\% &  5.5\% & 272.7 & 0.779 & 10{,}690 \\
L1: Single Evo. & 29{,}636 & 11.3\% & 12.6\% & 179.2 & 0.807 & 17{,}087 \\
L2: Stratified  & 37{,}960 & 14.5\% & 13.4\% & 342.8 & 0.836 & 28{,}112 \\
\bottomrule
\end{tabular}
\end{adjustbox}
\endgroup
\end{table}

\paragraph{Quantum vs.\ classical sampling}
Having established that the advantage originates at the sampling stage and that the stratified strategy is effective, we compare QSAD directly against three classical sampling baselines: uniform random sampling, greedy local search, and simulated annealing (SA). All methods operate on the same tetrahedral lattice with the same Hamiltonian under a matched pool-size protocol. For each protein, Random generates a pool of the same size as QSAD; Greedy and SA run until their pools reach the same size (within $\pm 20\%$), with a cap of 100{,}000 energy evaluations.

\begin{figure}[htbp]
\centering
\includegraphics[width=\linewidth]{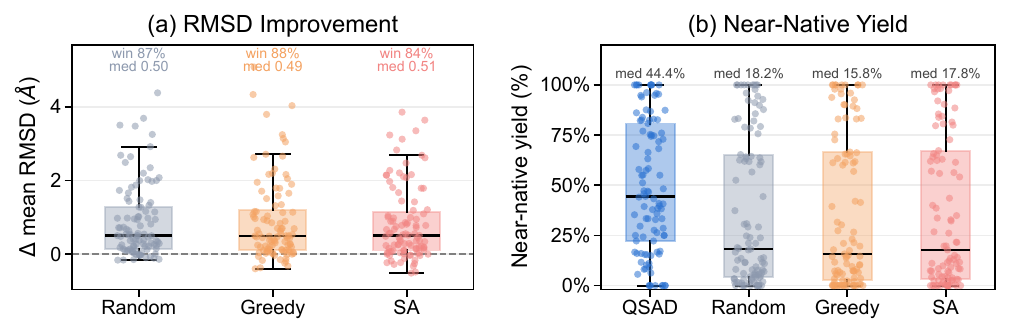}
\caption{Candidate pool quality under matched pool-size protocol. (a)~Pairwise improvement in mean valid RMSD over each classical sampler; positive values indicate QSAD wins. (b)~Near-native fraction (RMSD $< 4$~\AA{}): QSAD median 44\% vs.\ 16--18\% for all classical baselines.}
\label{fig:pool-quality}
\end{figure}

Figure~\ref{fig:pool-quality} compares candidate pool quality. QSAD wins on 87\% of proteins against Random, 88\% against Greedy, and 84\% against SA in mean valid RMSD. The near-native yield (RMSD $< 4$~\AA{}) shows a larger gap: QSAD achieves a median near-native fraction of 44\%, compared with 18\% for Random, 16\% for Greedy, and 18\% for SA. The advantage does not come from generating more valid conformations; rather, QSAD's valid conformations concentrate closer to the native structure.

\begin{figure}[htbp]
\centering
\includegraphics[width=\linewidth]{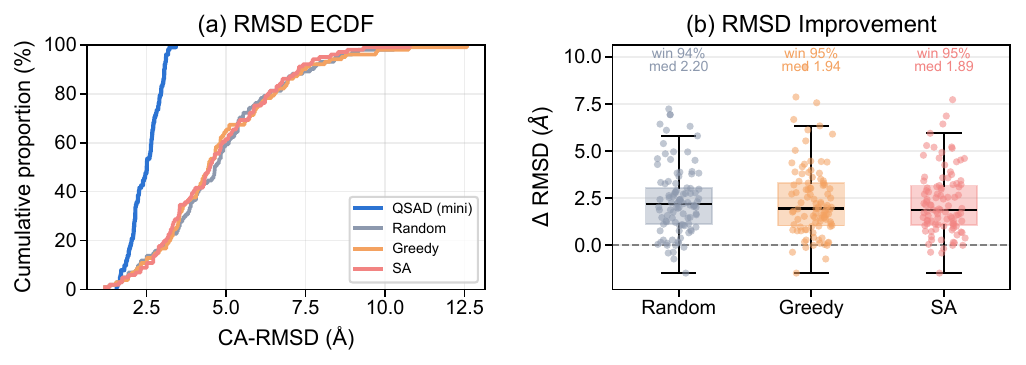}
\caption{End-to-end RMSD after energy-based selection and reconstruction. (a)~Cumulative distribution: QSAD median 2.51~\AA{} vs.\ 4.44--4.65~\AA{} for classical baselines. (b)~Pairwise improvement: QSAD wins on 94--95\% of proteins.}
\label{fig:end-to-end}
\end{figure}

After selecting the lowest-energy valid conformation from each pool and reconstructing it through the same all-atom pipeline, QSAD achieves a median end-to-end RMSD of 2.51~\AA{}, compared with 4.65~\AA{} for Random, 4.44~\AA{} for Greedy, and 4.49~\AA{} for SA (Figure~\ref{fig:end-to-end}). QSAD produces the lower-RMSD structure on 94\% of proteins against Random, 95\% against Greedy, and 95\% against SA, with median pairwise improvements of 2.20, 1.94, and 1.89~\AA{} respectively.

\section{Discussion}
\label{sec:discussion}

\paragraph{Why physics-driven sampling works for short peptides.}
The results show that the two requirements identified in the introduction must be satisfied together in this regime. The 44\% improvement over AlphaFold3 does not come from a larger model or more training data, but from encoding folding physics directly into the Hamiltonian and sampling that Hamiltonian effectively. The quantum versus classical comparison in Section~\ref{sec:system_validation} shows that the advantage lies in sampling quality: Hamiltonian evolution concentrates samples near low-energy basins more effectively than random, greedy, or simulated annealing search on the same Hamiltonian, winning on 95--96 out of 101 proteins. The primary quantum contribution is therefore the structure of the sampling distribution, not only faster execution than iterative methods.

\paragraph{Encoding trade-off and precision ceiling.}
The central design choice in QSAD is amino-acid-level encoding on a tetrahedral lattice. This keeps qubit cost at $O(N^2)$ and makes execution feasible on current hardware. The trade-off is a precision ceiling: lattice discretization limits the best achievable RMSD to about 1.6~\AA{}, regardless of sampling quality. A finer lattice or continuous coordinate encoding could raise this ceiling, but only at the cost of substantially higher qubit demand. In the current NISQ regime, this trade-off is appropriate because qubit budget remains the main constraint.

\paragraph{Limitations and scope.}
QSAD still has important limitations. The $\beta$ schedule ${1,2,3,4}$ is chosen heuristically rather than optimized per protein. The reconstructed landscapes are 2D PCA projections and should be viewed as approximate structural energy maps, not full thermodynamic free-energy surfaces, although Boltzmann-consistency checks support their utility.

\section{Conclusion}

QSAD shows that current quantum hardware, when paired with an appropriate problem formulation, can solve a meaningful computational biology task with strong accuracy, robustness, and execution efficiency. The key design choice is to encode folding physics at the amino-acid level rather than the atomic orbital level. This keeps qubit counts within the reach of existing processors while preserving enough physical structure to outperform all evaluated AI and quantum baselines on binding-pocket peptide prediction. More broadly, the results establish amino-acid-level, non-iterative quantum sampling as a practical computational path for short peptides in regimes where data-driven methods lack sufficient signal.

\bibliographystyle{unsrt}
\bibliography{main}

\end{document}